\documentclass[11pt]{article}
\usepackage{mons}
\usepackage{graphicx}
\usepackage{times}

\newcommand{\MONS}{{\em MONS\/}}

\lefthead{T. R. Bedding}
\righthead{Pulsating red giants}
\begin{document}

\title{Prospects for observing pulsating red giants with the \MONS{} Star
Trackers%
\enlargethispage{7ex}
\footnote{To appear in {\em Proceedings of the Third MONS Workshop: Science
Preparation and Target Selection}, edited by T.C.V.S. Teixeira and
T.R.~Bedding (Aarhus: Aarhus Universitet).}
}

\author{Timothy~R.~Bedding}
\affil{School of Physics, University of Sydney 2006, Australia}

\begin{abstract}

K and M giants show variability on timescales from years down to days and
possibly even hours.  I discuss the contribution that can be made with
high-precision photometry that will be obtained by the \MONS{} Star
Trackers.  These include observations of flare-like events on Mira
variables, and oscillation spectra for K~giants and short-period M~giants.

\end{abstract}

\section{Introduction}

The Star Trackers on the \MONS{} satellite (Bedding \& Kjeldsen, these
Proceedings) should produce exquisite light curves for many hundreds of red
giant stars.  These observations, made over about 30 days with high duty
cycle, will allow a number of questions to be addressed.  Classes of stars
are discussed in order of decreasing effective temperature, starting with
the Mira variables.

\section{Mira variables}

\begin{figure}
\centerline{ \includegraphics[width=10cm,bb=73 377 553
692]{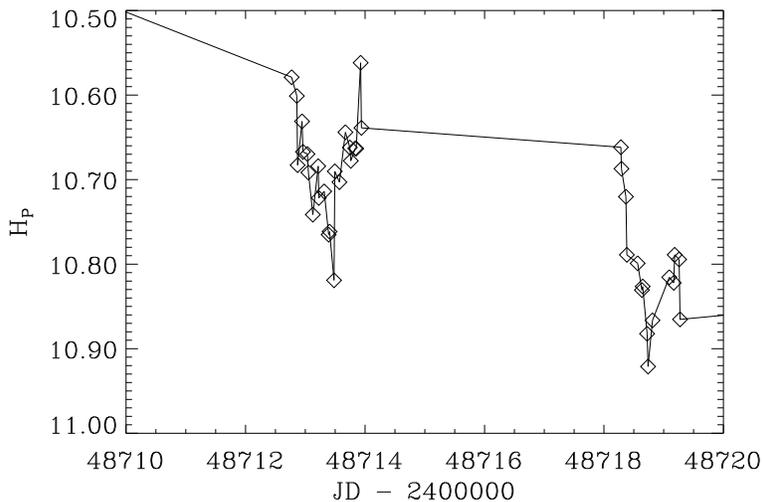} }
\caption[]{\label{fig.xhya} {\em Hipparcos\/} photometry of the Mira
variable HIP~47066 (X~Hya), showing two dips each lasting about one day. }
\end{figure}

Miras have the largest amplitudes and longest periods of all pulsating
stars.  What can we then hope to learn by studying them at high precision
over a month, which is only a small fraction of a pulsation cycle?
\cite{Scha91} has collected 14 cases of flares reported on Miras, lasting
minutes to hours and having amplitudes up to a magnitude.  A systematic
study by \cite{M+T95} found short-term variations in the photographic light
curves 18 long-period variables in M~16, with amplitudes $\ge0.5$
magnitudes and durations 1--30 days.  Most recently, \cite{deLMM98}
detected variations from {\em Hipparcos\/} photometry of Miras, with
amplitudes 0.2 to 1.1 magnitudes and durations from 2 hours up to 6 days.
In some cases, repeat events were observed on the same star (see
Fig.~\ref{fig.xhya}).

The origin of these short-term variations is not clear, but they are
presumably due to rapid and probably localized temperature changes.  One
possible cause might be the arrival at the surface of an unusually large
convection cell.  Given the high precision of the \MONS{} Star Trackers, we
should expect to see a distribution of events down to much smaller
amplitudes.  Two-colour information would be especially useful for shedding
light on this phenomenon.

\section{Short-period M giants}

\cite{K+L2000} (2000) have studied what they describe as rapidly
oscillating M giant stars.  They present a few dozen M giants that were
discovered by {\em Hipparcos\/} to have periods shorter than 10 days and
amplitudes up to a few tenths of a magnitude.  The only viable explanation
seems to be pulsation in very high overtones, and some stars shows signs of
multiple periodicities.  \cite{K+L2000} list about 35 stars with periods
less than 10 days, having $V$ magnitudes from 5.4 to 8.9 and amplitudes
mostly below 0.1\,mag.  Several of these stars will be observed by the
\MONS{} Star Trackers, and the light curves should allow a proper frequency
analysis for multiple modes.

\section{Oscillations in K giants}

It has been established from ground-based photometry that variability in
red giants decreases in amplitude as one moves down the spectral sequence
from M to K (\cite{JMS97}, \cite{FHH2000}).  With the kappa mechanism no
longer functioning, excitation is presumably due to convection.  Periods
become shorter as stellar density increases, and variablility becomes less
regular, presumably due to the stochastic nature of the excitation process
and/or to the presence of multiple modes.

A few bright K giants show radial-velocity (RV) variations that could be
due to oscillations, but it has proved difficult to obtain time series that
are long and continuous enough to resolve the frequency spectrum.  Matters
are complicated by the presence of long-term variations (hundreds of days),
which could be due to pulsation, rotational modulation or low-mass
companions (e.g., \cite{H+C93}, 1999).

The best-studied example is Arcturus ($\alpha$~Boo), which has been found
to have short-term RV variations with periods of a few days
(\cite{SMM87}, \cite{BJP90}, \cite{H+C94a}, \cite{Mer99}), as well as
long-term variations with a 
period of a few hundred days (\cite{H+C93}).  This star will definitely be
observed by the \MONS{} Star Trackers, since it lies close to $\eta$~Boo, a
high-priority primary target.  It is not clear whether 30-50\,d will be
long enough to produce a usable oscillation spectrum, but nearly-continuous
coverage over this period will produce a time series far better than any RV
observations so far obtained.

\begin{figure}
\centerline{ \includegraphics[width=\the\hsize,bb=65 59 555 617]
{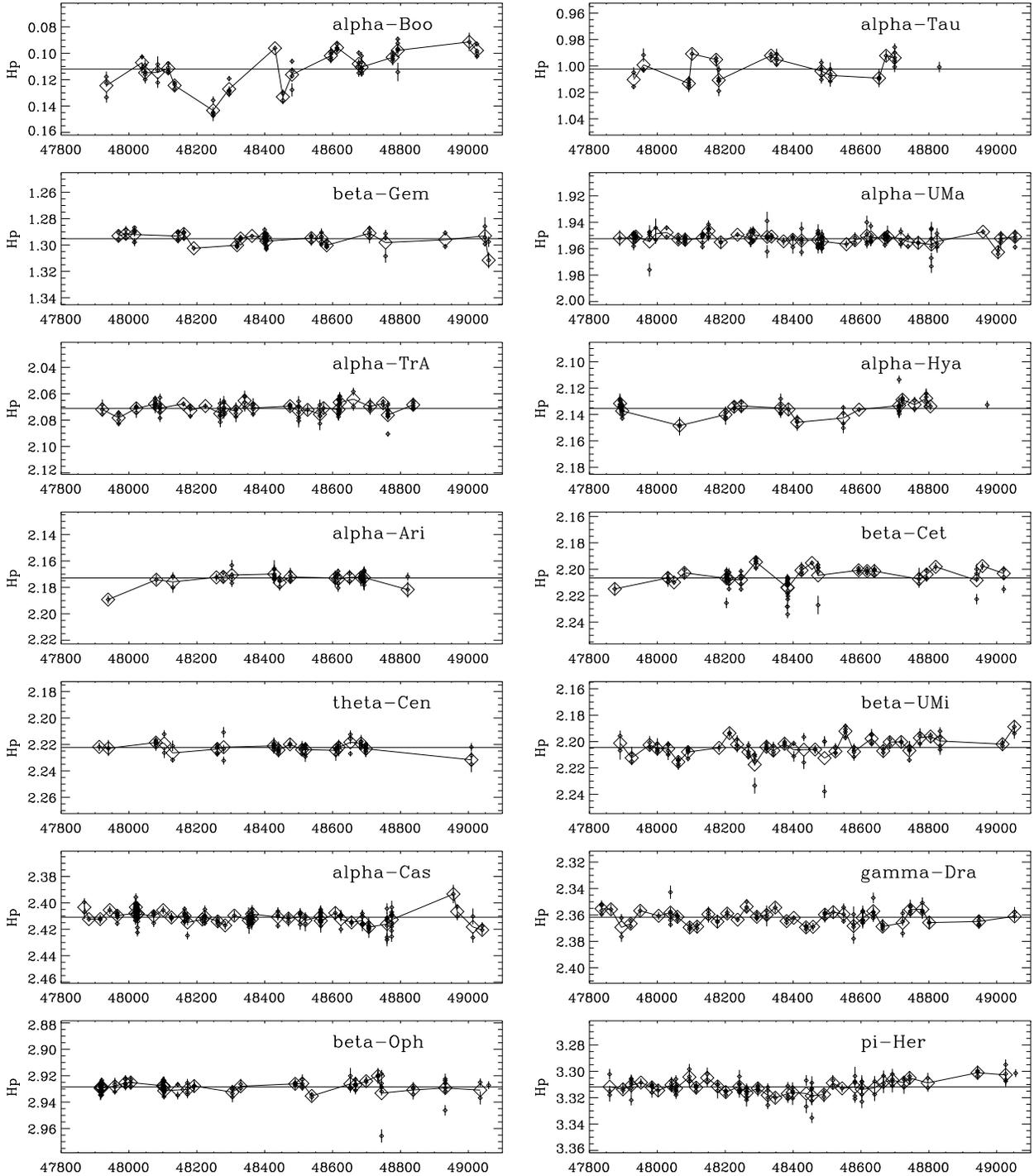} }
\caption[]{\label{fig.kgiants} {\em Hipparcos\/} photometry of 14 bright K
giants.  Small diamonds with error bars are individual measurements, while
larger diamonds (connected by a line) are means in one-day bins.  The
horizontal axis shows JD minus 2400000, and the vertical axis in each case
has a range of 0.1 magnitudes, with the horizontal lines showing the means
of the observations. }
\end{figure}

Figure~\ref{fig.kgiants} shows {\em Hipparcos\/} light curves for a sample
of the brightest K~giants (many of which are also known to be RV
variables).  Suprisingly, these light curves have not yet been discussed in
the literature.  We see clear evidence for photometric variability in
several stars, and we can confirm that Arcturus is indeed a variable star,
with peak-to-peak variations of about 0.04\,mag.

Another interesting case is $\pi$~Her, for which \cite{H+C99} obtained
RV measurements over two years that showed variability with a period of
about 600\,d.  They pointed out that if rotational modulation of surface
structure were the cause, one would expect photometric variations of about
0.1\,mag (peak-to-peak).  The {\em Hipparcos\/} light curve was obtained at
roughly the same time and shows some evidence for slow variability, but at
level about ten times smaller than this, allowing us to rule out spots as
the cause of the RV variations in $\pi$~Her.

Photometric variability in K~giants has previously been seen in globular
clusters.  \cite{E+G96} observed 47~Tuc with the {\em Hubble Space
Telescope\/} over 38.5\,hr and found variables with periods of 2--4 days
and semi-amplitudes of 5--15\,mmag.  \cite{KKS98} detected 15 red
variables in 47~Tuc from the Optical Gravitational Lensing Experiment
(OGLE), which had poorer precision but much better temporal coverage.
Their K-giant variables have periods of 2--36\,d and semi-amplitudes of
40--90\,mmag.  Interestingly, both these observations are predated by a
report by \cite{Yao90} of a red giant variable in the globular cluster
M~15, with a period of 4.3\,hr and an amplitude of about 20\,mmag.

It seems clear that many K giants are variable on timescales of hours to
days, and observations with the \MONS{} Star Trackers should produce
excellent light curves.  

Finally, we mention the recent exciting results by \cite{BCL2000}
(2000 and these Proceedings), who used the star camera on the
failed {\em WIRE\/} satellite to perform high-precision photometry of the
bright K~giant $\alpha$~UMa.  They produced evidence for multi-mode
oscillations in this star with periods of 0.3--6\,d and amplitudes of
0.1--0.4\,mmag.  Data of much higher quality are expected from the \MONS{}
Star Trackers and should produce rich oscillation spectra for a sample of
bright K~giants.

\subsection*{Acknowledgments}

This work was supported by the Australian Research Council.

\end{document}